# Angle of Arrival Detection with Fifth Order Phase Operators

Youssef Khmou, Said Safi

*Abstract*—In this paper, a fifth order propagator operators are proposed for estimating the Angles Of Arrival (AOA) of narrowband electromagnetic waves impinging on antenna array when its number of sensors is larger than the number of radiating sources.
The array response matrix is partitioned into five linearly dependent phases to construct the noise projector using five different propagators from non diagonal blocks of the spectral matrice of the received data; hence, five different estimators are proposed to estimate the angles of the sources. The simulation results proved the performance of the proposed estimators in the presence of white noise comparatively to high resolution eigen based spectra.

## I. INTRODUCTION

RETREIVING the information from the received signals by an antenna is very useful in many engineering fields, indeed estimating the Direction Of Arrival (DOA) of narrowband radiating sources has many applications [1] including synthetic aperture Radar, geo-location systems, submarine acoustic and chemical sensor array.

Most of the DOA estimators rely on second order statistics of the received data; they are divided into parametric and non parametric methods. The beam forming techniques [1] have the advantage of light computational load with possibility to estimate the powers of the received signals but provide a low resolution that depends on the geometry of the array. The subspace based techniques [2], [3] require the eigen decomposition or the singular value decomposition of the cross correlation matrix of the received signals to construct two orthogonal sets which are the signal and noise subspaces, after the estimation of the noise subspace, the angular spectrum is obtained by testing the angles in the range that is dependent on the geometry of the array, when the tested angle is DOA the spectrum represents a peak.

In some cases, it is required to use a large number of sensors to estimate the characteristics of the radiating sources, using the subspace based techniques is computationally expensive especially when the equipment used does not possess sufficient Random Access Memory (RAM). To get faster solutions, the Propagator operator [4], [5], [10] is proposed, to avoid the eigen decomposition, which exploits the linear dependence between the rows of the array matrix such that the noise subspace can be estimated with lower computational complexity. A new versions of the propagator were proposed [6], [7], using partial cross correlation sub-matrices, which are effective even in the presence of spatially and temporally non uniform noise.

Most of the DOA estimators require that the number of the sensors must be strictly superior to the number of emitters, but this condition can be surpassed using KHATRI RAO subspace [9] when the signals of the sources are quasi stationary which permits the construction of an augmented cross correlation matrix. However in some applications the number of the receivers can be much greater than the number of sources, which is the case in submarine acoustic [8], radiolocation [7] and also the new conception of antennas consisting of hundreds of cellular sensors.

In this paper, we introduce the fifth order propagator, a linear operator that can efficiently constructs the noise subspace of the narrowband model when the number of sensors is superior to four times the number of sources.

## II. STATISTICAL DATA MODEL

We consider a Uniform Linear Array (ULA) consisting of N identical sensors, the narrowband wave field created by P sources, located at distinct azimuth angles, is received by the antenna with N >> P. The property of isotropic medium allows us to express the received signals at the instant $t_i$ by the following equation:

$$X(t_i) = \sum_{k=1}^{P} a(\theta_k) s_k(t_i) + n(t_i) = A(\theta) s(t_i) + n(t_i) \quad (1)$$

The discrete time index $t_i = 1,2,\dots,K$ with $K$ is the number of acquired samples and $X(t) \in C^{N \times K}$ is the random matrix of the received signals, $A(\theta) = [a(\theta_1), a(\theta_2), \dots, a(\theta_P)] \in C^{N \times P}$ is the array steering matrix, $s^T(t) = [s_1(t), \dots s_P(t)]$ is the matrix of the sources waveforms.

$n(t) \in C^{N \times K}$ is the additive noise of the antenna sensors that is modeled by stationary and ergodic zero mean complex valued random process, whose joint probability density function is given by :

$$p(n(t)) = \frac{1}{\pi^N |\Gamma_n|^{1/2}} \exp\{\frac{n^+(t)\Gamma_n^{-1}n(t)}{2}\} \quad (2)$$

where $\Gamma_n = \sigma^2 I_N$ is the noise correlation matrix and $(.)^+$ denotes the conjugate transpose operator with $I_N$ being the identity matrix. The Vandermonde steering vector is defined by:

$$a(\theta_i) = [1, e^{-j\mu_i}, \dots, e^{-j(N-1)\mu_i}]^T$$

Youssef Khmou and Said Safi are with Department of Mathematics and Informatics, polydisciplinary faculty, Sultan Moulay Slimane University, Beni Mellal, Morocco (e-mail: khmou.y@gmail.com, safi.said@gmail.com).

with the path difference $\mu_i = 2\pi d \lambda^{-1} \sin(\theta_i)$ where $\theta_i \in \left[\frac{-\pi}{2}, \frac{\pi}{2}\right]$ is the $i^{th}$ Direction Of Arrival (DOA), $\lambda = c/f_c$ is the wavelength of the narrowband sources, c is the speed of electromagnetic wave and $f_c$ is the carrier frequency. d is the inter-element distance of the antenna and $(.)^T$ denotes the matrix transposition.

The spectral matrix $\Gamma$ can be computed in the time or frequency domain as follows:

$$\Gamma = E\{X(t)X^+(t)\} = A(\theta)\Gamma_s A^+(\theta) + \Gamma_n \quad (3)$$

$E\{.\}$ denotes the expectation operator, the matrix $\Gamma_s = E\{s(t)s^+(t)\}$ is the $P \times P$ complex covariance of the sources waveforms.

The subspace based techniques split the matrix $\Gamma$ into two orthogonal sets called signal and noise subspaces $(U_s, U_n)$ characterized by the property:

$$U_s U_s^+ + U_n U_n^+ = I_N \quad (4)$$

In the next section we introduce a new DOA estimator which exploits all elements of the antenna by using small blocks of the matrix $\Gamma$ to construct the set $U_n$.

### III. Proposed Operators

If the number of sensors is superior to four time the number of sensors $N \gg P$, we can divide the array into five non overlapping subarrays given that the radiating sources are statistically independent such that $\Gamma_s$ can be written as $\Gamma_s = diag\{\sigma_{s1}, \sigma_{s2}, \ldots, \sigma_{sP}\}$ or at least $\Gamma_s$ is invertible with $\sigma_{si}$ denotes the signal power of the $i^{th}$ source. Based on these assumptions, we can derive the fifth order propagator as follows: We take the first element of the array as the phase reference of the incoming signals, the steering matrix is given by:

$$A(\theta) = \begin{pmatrix} 1 & 1 & \ldots & 1 \\ e^{-j\mu_1} & e^{-j\mu_2} & \ldots & e^{-j\mu_P} \\ \ldots & \ldots & \ldots & \ldots \\ e^{-j(N-1)\mu_1} & e^{-j(N-1)\mu_2} & \ldots & e^{-j(N-1)\mu_P} \end{pmatrix} \quad (5)$$

where $A(\theta)$ has the properties $rank\{A(\theta)\} = P$, $A(\theta)°A^*(\theta) = 1_{N \times P}$. The following partitioning is generated:

$$A(\theta) = (A_1^T, A_2^T, A_3^T, A_4^T, A_5^T)^T \quad (6)$$

where $A_i \in C^{P \times P}$ for $i = 1,\ldots,4$ and $A_5 \in C^{N-4P \times P}$. From (5) we realize that any sub matrix $A_i$ is linearly dependent to $A_j$, hence there exist an ensemble $\Omega$ given by :

$$\Omega = \{\Pi_{ji} : C^{P \times P} \rightarrow C^{N-4P \times P} (C^{N-4P \times P} \rightarrow C^{P \times P}) : A_j = \Pi_{ji} A_i, (i,j) = 1,\ldots 5\}$$

Next, we search for a combination of operators $\Pi_{ji}$ to construct the null space of the steering matrix $A(\theta)$. For this purpose, $\Pi_{ji}$ can be extracted from the cross correlation matrix $\Gamma$ such that for any indexes $(i,j)$, the block $\Gamma_{ij}$ is given by:

$$\Gamma_{ij} = f(A_i, A_j) = A_i \left\{\frac{1}{K}\sum_{t=1}^{K} s(t)s^+(t)\right\} A_j^+ \quad (7)$$

with the condition $i \neq j$ to eliminate the affected blocks by the perturbation noise $n(t)$ as illustrated in the following matrix :

$$\Gamma = \begin{pmatrix} \Gamma_{11} + \sigma^2 I_P & \Gamma_{12} & \ldots & \ldots & \Gamma_{15} \\ \Gamma_{21} & \Gamma_{22} + \sigma^2 I_P & \ldots & \ldots & \Gamma_{25} \\ \ldots & \ldots & \ldots & \ldots & \ldots \\ \ldots & \ldots & \ldots & \ldots & \ldots \\ \Gamma_{51} & \ldots & \ldots & \ldots & \Gamma_{55} + \sigma^2 I_{N-4P} \end{pmatrix}$$

Any phase $A_j$ can be expressed as a function of $A_i$ with the following indicial product:

$$A_j = \Gamma_{jk}\Gamma_{ik}^{-1}A_i = A_j\Gamma_s A_k^+ A_k^{+-1}\Gamma_s^{-1} A_i^{-1} A_i = \Pi_{ji} A_i \quad (8)$$

The equation is valid for any $k \neq (i,j)$, after the construction of the ensemble $\Omega$, we can calculate the fifth order propagator that approximate the null space of $A(\theta)$ based on any partition in (6). Per example we take $A_3$ to be the base of equations, we have:

$$A_3 = \Pi_{31}A_1 = \Pi_{32}A_2 = \Pi_{34}A_5 = \Pi_{35}A_5 \quad (9)$$

It follows that the propagator operator $\Psi_{53}$ is computed by:

$$\Psi_{53} A(\theta) = [\Pi_{31}|\Pi_{32}| - 4I_P|\Pi_{34}|\Pi_{35}] A(\theta) = 0_{P \times P} \quad (10)$$

Similarly to the remaining phases $A_i$, we compute the five different operators that can be concatenated in single matrix $\Psi \in C^{N \times N}$ as follows :

$$\Psi = \begin{pmatrix} \beta_P & \Pi_{12} & \Pi_{13} & \Pi_{14} & \Pi_{15} \\ \Pi_{21} & \beta_P & \ldots & \ldots & \ldots \\ \Pi_{31} & \ldots & \beta_P & \ldots & \ldots \\ \Pi_{41} & \ldots & \ldots & \beta_P & \ldots \\ \Pi_{51} & \ldots & \ldots & \ldots & \beta_{N-4P} \end{pmatrix} \quad (11)$$

where $\beta_n = -4I_n$, the operators $\Psi_{5i}$ have the same resolution capability. In the absence of the perturbation (i.e $X(t) = A(\theta)s(t)$) we write the following equations:

$$\varphi(\Psi_{5i}, A(\theta)) = \frac{\pi}{2} \quad (12)$$

$$\Psi \Gamma = 0_{N \times N} \quad (13)$$

$$\Psi X(t) = 0_{N \times K} \quad (14)$$

The eigenvalues and singular values of $\Psi$ have a complementary property compared to that of the spectral matrix $\Gamma$, theirs spectra are given by :

$$\sigma_\Gamma = \{\lambda_1 \geq \lambda_2 \geq \cdots \geq \lambda_P > \lambda_{P+1} = \cdots = \lambda_N = 0\} \quad (15)$$

$$\sigma_\Psi = \{\lambda_1 = \lambda_2 = \cdots = \lambda_P = 0 < |\lambda_{P+1}| \leq \cdots \leq |\lambda_N|\} \quad (16)$$

From (11) and (16) we deduce that:

$$rank\{\Psi\} = N - P \quad (17)$$

$$Trace\{\Psi\} = -4N \quad (18)$$

The angular spectrum for the $i^{th}$ operator is given by the following metric:

$$f_i(\theta) = \frac{a^+(\theta)a(\theta)}{a^+(\theta)\Psi_{5i}^+\Psi_{5i}a(\theta)} \quad (19)$$

In the next section we give some simulation results to evaluate the performance of the proposed operators.

## IV. SIMULATION RESULTS

We consider an array consisting of N=18 identical and isotropic sensors, P=3 non coherent narrowband sources are impinging on the array with the same power of unity. The number of samples is set to K=200. The carrier frequency is $f_c = 1 GHz$ and the sensors are equally spaced with a distance of half the wavelength d=15cm with Rayleigh limit angular resolution of $\theta_{HPBW} \cong 6.74°$. The signals of the sources are assumed to be complex ergodic random processes. In the first experiment we set $SNR = 5dB$, Fig. 1 represents an average of $L = 100$ Monte Carlo runs of the five operators.

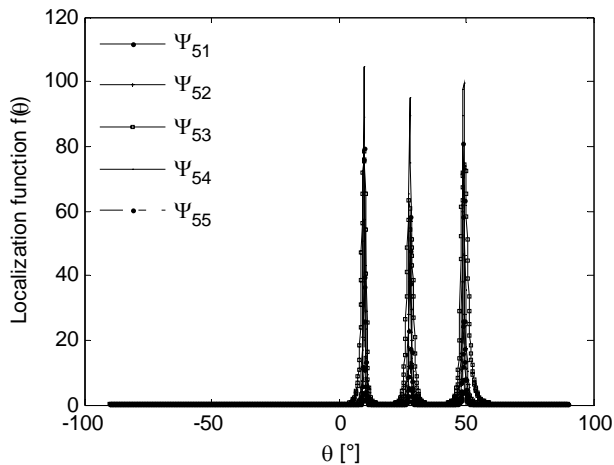

Fig. 1 Proposed operators with $N = 18, P = 3$, $\theta = [10°, 28°, 49°], d = \lambda/2, K = 200$ and SNR = 5dB

We realize that all operators have the same behavior with no side lobes which make the physical interpretation easy. To compare the proposed operators with some high resolution spectral techniques, we evaluate the operator $\Psi_{55}$ with Schmidt subspace based technique [2] and standard Beam forming, the results of $L = 100$ Monte Carlo runs are presented in the Fig. 2.

The operator $\Psi_{55}$ and Music are equivalent in these conditions. In the last experiment we compare the Root Mean Square Error of the operators $\Psi_{5i}$ with high resolution ESPRIT Technique [3] (Estimation Of Signal Parameters Via Rotational Invariance Techniques) the result, of $L = 100$ Monte Carlo for every value of SNR, is presented in Fig. 3.

We conclude in this experiment that the operators $(\Psi_{55}, \Psi_{51})$ are better performing than the three other functions and when the SNR is about 20dB they are equivalent to ESPRIT.

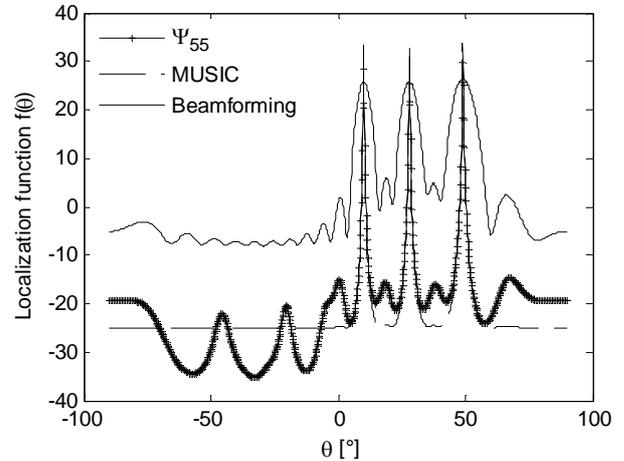

Fig. 2 Spectra $\Psi_{55}$, MUSIC and Beamforming with $N = 18, P = 3, \theta = [10°, 28°, 49°], d = \lambda/2, K = 200$ and SNR = 5dB

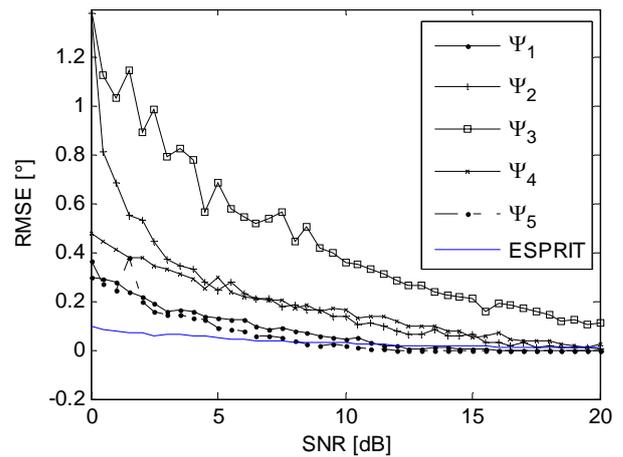

Fig. 3 RMSE against SNR=[0dB,20dB] over 100 Monte Carlo runs with $N = 18, P = 3, \theta = [10°, 28°, 49°], d = \lambda/2$ and $K = 200$

If the source waveforms are correlated, then forward backward averaging techniques are necessary for the proposed estimators to efficiently resolve the angles. In the other hand, the only drawback of the proposed approach is that the number of sources must be known; therefore we need to elaborate non eigen-based spectral estimators to detect the number P.

## V. CONCLUSION

We proposed, in this paper, the fifth order phase operators for narrowband DOA estimation which is applicable when the number of sensors in the antenna is larger than the number of radiating sources.

The proposed idea consists of dividing the array steering matrix into five linearly dependent phase from which we derived five almost equivalent noise subspace projectors, where each operator is computed by exploiting the linear maps between the corresponding blocks of the spectral matrix. The simulation results confirmed the validity of the proposed approach comparatively to standard high resolution spectral techniques.